\documentclass{vgtc}                     

\newif\ifnotes
\notesfalse 

\newif\iftrack
\trackfalse 

\onlineid{0}

\vgtccategory{Research}

\vgtcpapertype{application/design study}

\usepackage{microtype}                 
\PassOptionsToPackage{warn}{textcomp}  
\usepackage{textcomp}                  
\usepackage{mathptmx}                  
\usepackage{times}                     
\usepackage{ccicons}
\usepackage{booktabs}                  
\usepackage{float}
\usepackage{dblfloatfix} 
\usepackage{lettrine}
\usepackage{csquotes}

\usepackage[normalem]{ulem}

\usepackage{enumitem}
\usepackage{fontawesome5}
\usepackage[linesnumbered,ruled,vlined]{algorithm2e} 
\usepackage{algorithmicx}
\usepackage{algpseudocode}
\usepackage{amsmath}

\usepackage{xargs} 
\usepackage{soul}  
\usepackage{color} 
\usepackage{xspace}
\usepackage{listings}

\newcommand{\remixtape}{\textsc{RemixTape}}
\newcommand{\papertitle}{\remixtape: Enriching Narratives about Metrics with\\Semantic Alignment and Contextual Recommendation}

\definecolor{forestgreen}{RGB}{50,120,50}
\definecolor{rtblue}{RGB}{74,150,178}
\definecolor{rtgreen}{RGB}{2,177,127}
\definecolor{gray}{RGB}{167,167,167}

\newcommand{\metric}[1]{\textcolor{rtgreen}{{\small{\faHashtag}}~#1}}
\newcommand{\vizcard}{\texttt{VizCard}}
\newcommand{\vizcards}{\texttt{VizCard}s}
\newcommand{\textcard}{\texttt{TextCard}}
\newcommand{\textcards}{\texttt{TextCard}s}

\newcommand{\ie}{i.e.,\ }
\newcommand{\etal}{et al.}
\newcommand{\eg}{e.g.,\ }

\newcommand{\bstart}[1]{\vspace{1mm} \noindent{\textbf{#1.}}}
\newcommand{\bfstart}[1]{\vspace{1mm} \noindent{\textbf{#1}}}
\newcommand{\istart}[1]{\vspace{1mm} \noindent{\textit{#1.}}}
\newcommand{\bcstart}[1]{\vspace{1mm} \noindent{\textbf{#1:}}}

\newcommandx{\revision}[2][1=]{\iftrack{\textcolor{gray}{\st{#1}}\textcolor{red}{#2}}\else{#2}\fi}

\title{\papertitle}

\author{Matthew Brehmer\thanks{e-mail: mbrehmer@uwaterloo.ca}\\ %
    \scriptsize Unviersity of Waterloo %
\and Margaret Drouhard\thanks{e-mail: mar.drouhard@salesforce.com}\\ %
     \scriptsize Salesforce %
\and Arjun Srinivasan\thanks{e-mail: arjun.srinivasan@databricks.com}\\ %
     \scriptsize Databricks}

\abstract{%
    The temporal dynamics of quantitative metrics or key performance indicators (KPIs) are central to conversations in enterprise organizations. 
Recently, major business intelligence providers have introduced new infrastructure for defining, sharing, and monitoring metric values.
However, these values are often presented in isolation and appropriate context is seldom externalized.
In this design study, we present \remixtape, an application for constructing structured narratives around metrics.
With design imperatives grounded in prior work and a formative interview study, \remixtape~provides a hierarchical canvas for collecting and coordinating sequences of line chart representations of metrics, along with the ability to externalize situational context around them. 
\remixtape~includes affordances to semantically align and annotate juxtaposed charts and text, as well as recommendations of complementary charts based on metrics already present on the canvas. 
We evaluated \remixtape~in a study in which six enterprise data professionals reproduced and extended partial narratives. 
They appreciated \remixtape~as a novel alternative to dashboards, galleries, and slide presentations for supporting conversations about metrics.
\revision[We conclude with a reflection]{Finally, we reflect} on \revision[our design choices and process]{\remixtape's usability and potential utility}, \revision[]{and conclude} with a call to define a conceptual foundation for remixing in the context of visualization.

}


\teaser{
  \centering
  \includegraphics[width=\linewidth, alt={.}]{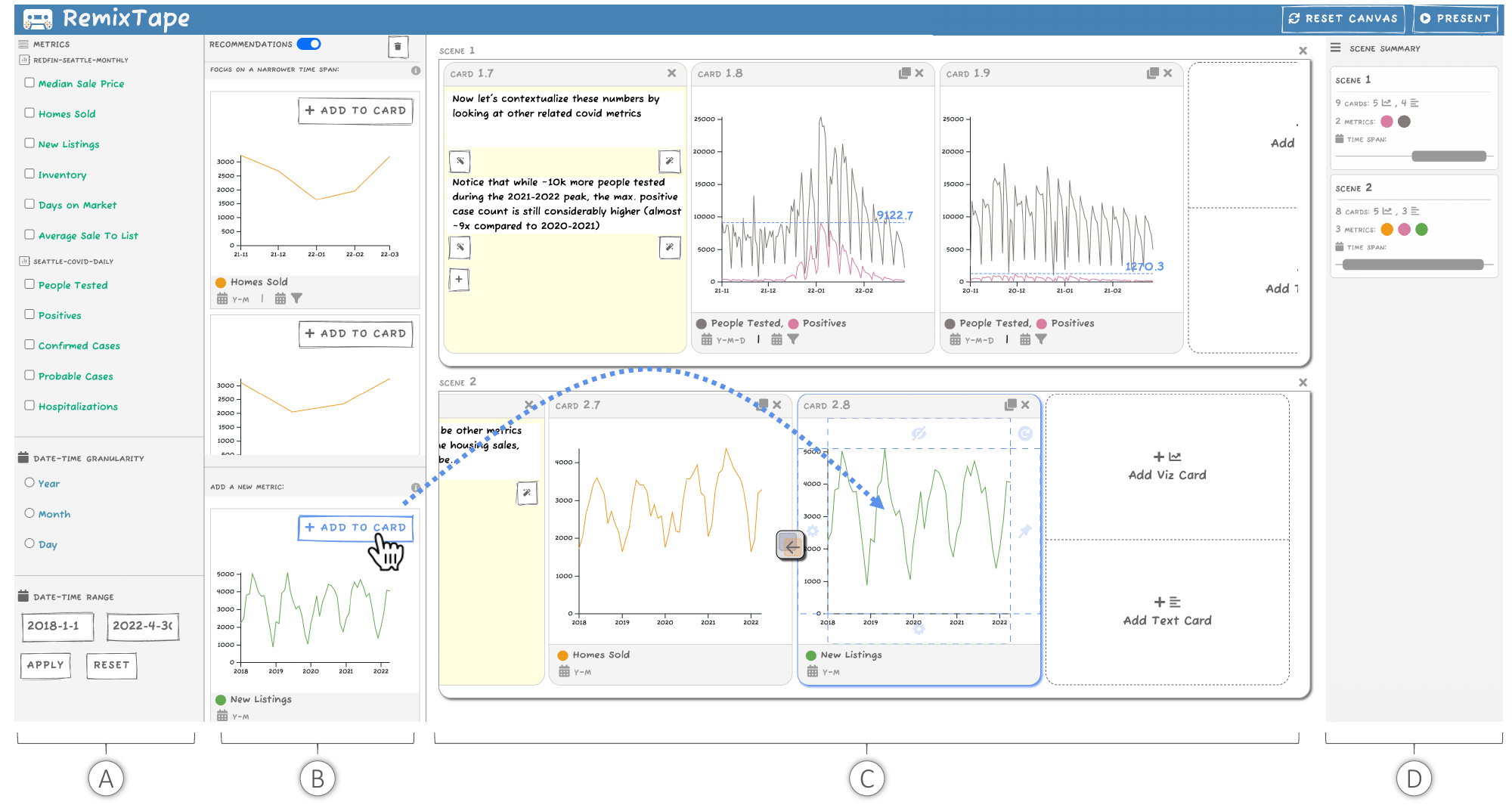}
  \caption{%
  	Upon adding a new \vizcard~(Card 2.8) to a scene in the authoring canvas (C), \remixtape~offers contextually-aware recommendations for populating the card (B) using available metrics (A). After selecting one of them (\metric{New Listings}), \remixtape~updates the scene summary (D) and reveals an affordance to semantically align this card with its neighbor (\metric{Homes Sold}).
  }
  \label{fig:teaser}
}

\graphicspath{{figs/}{figures/}{pictures/}{images/}{./}} 

\usepackage{tabu}                      
\usepackage{booktabs}                  
\usepackage{lipsum}                    
\usepackage{mwe}                       

\usepackage{hyperref}
\usepackage[capitalise]{cleveref}

\usepackage{mathptmx}                  

\begin{document}



\maketitle

\section{Introduction}\label{sec:introduction}

Conversations about data are ubiquitous in enterprise organizations~\cite{tory2021finding}.
Prior research~\cite{brehmer2021jam} highlights the prevalence of time-oriented data in these conversations with recurring references to quantitative \textit{metrics} (also commonly referred to as \textit{key performance indicators} or KPIs).
The temporal dynamics of metrics as depicted in line charts are central to the discussion of trends, the forming of predictions, and the setting of goals. 
The emergence of unified metrics platforms within enterprise business intelligence (BI) platforms~\cite{powerbimetrics,databrainmetrics,dbtmetrics,transformmetrics,tableaupulse} has made tracking and sharing metrics accessible to more people across an organization; metrics no longer need to be bound to dedicated analytical applications or individual dashboards.
Despite these new abilities, enterprise communication processes still require ways to emphasize the relationships between metrics and the context around them.

We present \remixtape, an application for enriching conversations around metrics.
\remixtape~affords coordinating and annotating line chart representations of metrics and arranging them into compelling narrative presentations (\cref{fig:teaser}). 
\remixtape~reconciles disparate temporal and quantitative value domains with affordances to semantically align juxtaposed metrics, thereby allowing metrics to contextualize one another.
It also provides context-driven recommendations of metrics that complement partially-constructed narratives.
Finally, it provides the means to externalize the situational context of metrics via annotation and text-chart synchronization.  

The \textbf{contributions} of this paper are threefold. 
First, we draw attention to BI practices involving arranging, coordinating, annotating, and re-contextualizing metrics, reflected in findings from a formative interview study and a synthesis of prior research.
Second, we describe \remixtape~as a novel combination of semantic alignment and context externalization with text-chart synchronization and chart recommendation.
Finally, we interpret findings from a study with six enterprise data professionals, leading to a critical reflection on \remixtape's \revision[plausibility]{interface design and plausibility.} 
\section{Background and Related Work}
\label{sec:rw}

We situate our research within the context of enterprise communication around metrics.
We then review prior work relevant to \remixtape's key elements: view coordination, context externalization, chart recommendation, and visualizing time-oriented data.

\subsection{Communicating Around Data in the Enterprise}
\label{sec:rw:communication}

Artifacts originating from a data analysis process may be combined and repurposed across an organization, either by the data analyst or data steward who originally created them, or by those in other roles~\cite{tory2021finding}.
Many of these assets are BI dashboards~\cite{sarikaya2018we}, and Tory~\etal~\cite{tory2021finding} characterized several related activities that take place to support the circulation of dashboard content, the most prevalent being \textit{`slicing and dicing'} dashboard content, imposing a narrative, changing the representation of the content, and ultimately sharing it.
Walchshofer~\etal~\cite{walchshofer2023transitioning} further observed that different specialists undertake data-related activities within an organization, this despite the reality that most BI dashboards are often designed with a single persona in mind. 
Lee~\etal~\cite{lee2015more} suggest that these unique roles include \textit{analysts}, \textit{editors}, and \textit{presenters}, meaning that those who prepare presentations about data may be distinct from those who constructed the charts and dashboards used in the presentation~\cite{brehmer2021jam}.
The aim of \remixtape~is to support the latter role: those who re-purpose and recombine data assets for the purpose of communication.

Many presentations about data taking place within enterprise settings~\cite{brehmer2021jam} manifest as what Segel and Heer~\cite{segel2010narrative} would characterize as \textit{slide shows} of annotated charts. 
These sequences~\cite{hullman2013deeper,hullman2017finding} of charts and text are often arranged in such a way so as to produce a particular rhetorical effect~\cite{hullman2011visualization}, to form an argument~\cite{kosara2017argument}, or to persuade the viewer to act~\cite{pandey2014persuasive}.
There is a considerable breadth of narrative design patterns for presenting these sequences~\cite{bach2018}, including those specific to presenting time-oriented data~\cite{lan2021understanding}, and we strived to ensure that this breadth of narrative structure is accessible with \remixtape.
Lastly, it is common practice~\cite{brehmer2021jam} to use slide presentation software such as Microsoft PowerPoint to present sequences of static chart images. 
In contrast, Ellipsis~\cite{satyanarayan2014authoring}, Timeline Storyteller~\cite{brehmer2019timeline}, and Tableau's Story Points~\cite{tableaustorypoints} have each demonstrated the capability to present sequences of interactive charts that can be dynamically augmented with annotations, a capability that we perpetuate with \remixtape. 

\remixtape~is also a response to the evolution of enterprise BI deployments, from dedicated applications and dashboards to headless BI and embedded experiences. 
Part of this shift is the adoption of unified metrics layers (or similarly-described offerings) from BI vendors, including Microsoft's Metrics for Power BI~\cite{powerbimetrics} and Tableau's Pulse~\cite{tableaupulse}.
While metrics can be monitored or shared individually, they can also be organized into folders, galleries, or collections~\cite{tableaucollections}. 
Multiple metrics can also be packaged as dashboards, such as in
Epperson~\etal's recent Quick Dashboard~\cite{epperson2023declarative}. 
The latter is a departure from the convention of bottom-up BI dashboard construction, in that metric definition and individual chart specification are automated, allowing for a top-down dashboard specification. 
Like Quick Dashboard, \remixtape~also obviates the need to define metrics or specify chart designs; it differs, however, in its reconciliation of metrics with different units and time periods, the arrangement of views into narrative sequences rather than dashboards, and its affordances for externalizing situational context. 

\subsection{Arranging and Coordinating Multiple Views}
\label{sec:rw:mv}

Central to \remixtape~is the arrangement of line chart representations of metrics through juxtaposition and superposition.
While a slideshow is one way to present a sequence of views~\cite{brehmer2019timeline,elias2013storytelling,gratzl2016visual,satyanarayan2014authoring}, there can also be value in considering approaches that distribute views over space, whether in a grid of orthogonally aligned views~\cite{gratzl2014domino}, in a loosely structured canvas of data comics~\cite{bach2018design}, or in a more structured dashboard~\cite{bach:2022dashboardDesignPattern,elias2013storytelling,deng2022revisiting,lin2022dminer,sarikaya2018we}. 
In each case, the relative size, placement, and salience of views direct a viewer's attention.
Another approach~\cite{conlen2018idyll,mckenna2017visual}, one exemplified in VizFlow~\cite{sultanum2021leveraging}, distributes views over both space and time in scrollable articles.
We distribute views in \remixtape~across a structured canvas of scenes, so a series of views can be visible concurrently. 

Regardless of whether a sequence of views are arranged across space, over time, or both, it is critical to coordinate these views~\cite{qu2017keeping} with consistent visual encoding choices, to clearly harmonize axes and units across adjacent views where appropriate, and to clearly indicate cases where view coordination is impossible or undesirable. 
It may also be possible to consolidate partially redundant juxtaposed views by superimposing them or by explicitly encoding the differences in their values~\cite{lyi2020comparative}. 
While we exclude the derivation of new metrics in \remixtape, as this is the responsibility of the data analyst or data steward, we incorporate functionality for harmonizing juxtaposed line charts as well as superimposing them when they are semantically related, functionality that can be described as semantic alignment. 
Kristiansen~\etal's characterization of \textit{semantic snapping}~\cite{kristiansen2021semantic} is a related concept, one that informs strategies to reduce redundancy and inconsistency across multi-view arrangements of charts, particularly in compact dashboards. 
However, in narrative presentations of data, compactness is less critical, and some degree of redundancy can be useful for emphasis and message retention.

\subsection{Externalizing Situational Context}
\label{sec:rw:annotating}

Many narrative visualization tools (\eg~\cite{brehmer2019timeline, elias2012annotating, satyanarayan2014authoring, tableaustorypoints}) allow for the annotation of charts, however it is worth unpacking what annotation entails.
Ren~\etal~\cite{ren2017chartaccent} describe a design space of annotation as having two dimensions: annotation \textit{forms} (including text, shape, highlight, and image) and annotation \textit{targets} (including data marks, coordinate spaces, prior annotations, and other elements such as axes or legends).
While tools like SmartCues~\cite{subramonyam2018smartcues} and ChartAccent~\cite{ren2017chartaccent} demonstrate coverage of this design space across different types of charts, we consider a subset of ephemeral and persistent annotations for line charts in \remixtape, which include reference line shapes for indicating goals or thresholds and text value annotations for drawing attention to specific peaks, troughs, or inflection points.

The ability to annotate charts with open-ended text commentary allows people to externalize situational context and prior knowledge.
Prior work demonstrates several points in the design space for this coordination, one that presents choices in such as the relative size and placement of text commentary and visualization elements as well as the level of interaction between these elements.
In a data analysis context, it can be useful to collect externalizations in a prominent dedicated view while retaining an interactive link to their associated chart referents~\cite{kim2019inking}.
In a narrative visualization context, one approach involves interleaving text and visualization within an article~\cite{conlen2018idyll}, where visualization elements could appear, update, or disappear in response to the scroll position~\cite{mckenna2017visual,sultanum2021leveraging}. 
Alternatively, Kori~\cite{latif2021kori} and DataTales~\cite{sultanum2023datatales} augment text with interactive widgets that trigger corresponding highlights or updates in adjacent charts.
Dashboards can blur the line between analytical and narrative experiences~\cite{Meeks2018}, particularly when they are populated with verbose and responsive captions containing inline filter and parameter controls, in what could be referred to as a \textit{narrative dashboard}~\cite{walchshofer2023transitioning}.
\remixtape~similarly places an equal level of emphasis on chart and text elements, and we coordinate them such that a text element can interactively highlight a spatially adjacent line chart.

\subsection{Visualization Recommendation}
\label{sec:rw:recommendation}

The need to re-purpose and \textit{slice and dice} BI assets~\cite{tory2021finding} implies a need to filter and align views from heterogeneous sources, as well as locate, browse, and identify relevant assets within shared repositories of dashboards, charts, datasets, and metrics.
Altogether, these activities may be supported or accelerated by system guidance in the form of contextual recommendations.
Prior research offers guidance strategies in visual analysis workflows~\cite{collins2018guidance,ceneda2019review,ceneda2016characterizing}, as well as approaches to visualization~\cite{lee2021deconstructing,wu2021survey,zhu2020survey} and dashboard design~\cite{lin2022dminer,pandey2022medley} recommendation.
Line chart recommendations in \remixtape~are intended to surface complementary metrics and of logical sequences of exposition for ensuring a coherent narrative.
Given this focus, we draw upon design ideas and techniques from visual data exploration recommendation systems that offer guidance by suggesting \textit{`next steps'} during data analysis (\eg\cite{gotz2009behavior,wongsuphasawat2015voyager,key2012vizdeck,wongsuphasawat2017voyager,bouali2016vizassist,siddiqui2016effortless}).
However, recommendations in \remixtape~fundamentally differ from prior systems in that they focus on contributing to a narrative sequence, as opposed to promoting data coverage~\cite{wongsuphasawat2015voyager} or supporting an open-ended data analysis process~\cite{lee2021deconstructing}.

\subsection{Visualizing Metrics over Time}
\label{sec:rw:time}

Although there are many ways to represent data over time~\cite{aigner2011visualization}, including those purpose-built for exploratory data analysis (\eg~\cite{zhao2011exploratory,zhao2011kronominer}) and monitoring (\eg\cite{mclachlan2008liverac,lin2004viztree}),  the line chart is an established and familiar visual idiom, particularly for communicating metrics in a BI context~\cite{powerbimetrics,tableaupulse}.
Apart from line charts, sparklines~\cite{tufte2006} and BANs~\cite{wexler2017} are also commonly used in BI contexts to show metric values compactly.
However, given their prevalence in practice, \remixtape~exclusively represents metrics as line charts.

We observe two special considerations for line charts in \remixtape. 
First, truncating the vertical axis of a line chart can be useful to draw a viewer's attention to value differences over time, though without sufficient context or annotation, this choice could be deceptive~\cite{correll2020truncating}; \remixtape~leaves the choice to truncate the vertical axis up to the discretion of the user.
A second consideration is whether to enable the creation of a dual-axis line chart, a convention used when one or more series of values exhibit a different value domain or unit of measure relative to what is already displayed in the chart. 
Following the empirical findings of Isenberg~\etal~\cite{isenberg2011study} indicating that dual axis charts can elicit erroneous value judgments, \remixtape~restricts authors from creating them, opting instead for the creation of \textit{indexed} line charts (\eg\cite{Bostock2018}), relativizing values to a common percentage scale indexed to an initial value.
\section{Formative Interview Study}
\label{sec:remixing}

To complement prior work in the BI domain (\cref{sec:rw:communication}), we conducted a series of semi-structured 60-minute interviews to better understand why and how people adapt and recombine BI assets in enterprise contexts.
We recruited six professionals who reported customizing, adapting, and repurposing BI assets to support sharing and storytelling across their organizations. 
The participants represented various sectors of the economy, including finance, higher education, technology, the automotive sector, and the nonprofit sector.

We began by asking participants about their role(s) and work related to BI assets: how they currently adapt, reuse, or share these assets, along with the challenges that arise when doing these activities. 
In the latter half of each interview, we used a dashboard containing sales metrics as a design probe and asked participants to explain how they might customize and annotate the content to share insights with others. 
Throughout the scenario, we asked participants about opportunities for future tool support.
The second author led the interviews and initial thematic analysis of the data collected findings, while the other authors assisted with a secondary focused analysis dedicated to the implications for metrics.

\bstart{Collecting and repurposing charts}
Interviewees described combining charts associated with different sources, such as disparate BI dashboards. 
However, this often entailed downloading raw data from a dashboard and painstakingly reproducing charts in spreadsheet and slide presentation tools in order to make necessary adaptations for a particular context; fittingly, a financial product sales analyst candidly remarked that this time-consuming duplication of work is \textit{`stupid'}.
A digital marketing producer in the higher education sector described the appeal of an \textit{``additional workspace where I can rearrange and move stuff,} [\ldots] \textit{pulling stuff from different dashboards and bring it all into one place.''}
However, the suggestion of such a workspace prompted a consultant from the automotive sector to insist that the original sources of charts should be clearly indicated.
Moreover, participants indicated that such a workspace should allow people to interact with and make changes to the collected charts, suggesting that static screenshots of charts and dashboards are insufficient. 
According to the consultant, existing processes for embedding live charts from different sources are \textit{``way too difficult.''}

\bstart{Arranging and coordinating charts}
The interviewer's suggestion of a whiteboard workspace for collecting charts from disparate sources resonated with all of the interviewees.
However, the free arrangement of charts across a digital whiteboard seemed ill-suited for some downstream activities, namely giving presentations and sharing reports, as these formats impose some degree of linear structure on the content they contain.
The automotive industry consultant suggested that \textit{``a storyboard kind of visual would help,''} as an alternative to a free-form whiteboard organizing structure.
A linear structure reaffirmed the need for annotation, as captured by a project manager at a non-profit agency: \textit{``I think just some commentary would be useful if you're presenting this} [content] \textit{to your team.''}

\bstart{Adding context and directing attention}
Assuming the possibility of collecting and consolidating charts within a common workspace, another aspect of customization described by interviewees involved adding annotations and enriching the charts with situational context.
For instance, the sales analyst described the appeal of free-form annotation tools available when using screen-sharing features in telecommunication tools (\eg Cisco WebEx Meetings~\cite{webex}), and that such annotation tools could also be useful in a workspace of collected charts: \textit{``you can use a little flashlight that highlights, and you can draw a box around something, you can draw lines along something, you can write your own text and basically annotate.''}  
Similarly, the digital marketing producer described a desire to add sticky notes to charts and dashboards, and she also described the ability to \textit{`mark up'} content and to toggle the visibility of this graphical annotation on demand.
A financial analyst at a technology company also reminded us of how individual annotations may have varying levels of scope: \textit{``annotations could mean data labels, data points, or it could mean simple one-line / two-line sentences.''}

\bstart{Focusing on metrics}
Participants described a recurring motivation to collect, arrange, and annotate charts depicting metrics, providing audiences with a unified context where they might compare and reason about values over time. 
For instance, the automotive industry consultant described that \textit{``a lot of clients want to look at the change and the difference,} [\ldots] \textit{KPI} [Key Performance Indicator] \textit{type of stuff.''}
Similarly, the digital marketing producer described contextualizing a metric by providing year-over-year value comparisons.

Beyond providing the means to draw comparisons between known metrics of interest, the project manager described the potential value of ad-hoc trend comparison once a set of metrics are gathered. 
In her words, there would be value in seeing an additional \textit{``trend or something that} [she] \textit{wouldn't even necessarily know to look for,''}
In other words, there is potential utility in surfacing recommendations that indicate relationships between metrics, recommendations that could complement a partially-constructed narrative.


\bstart{Design imperatives}
We distill our findings into design imperatives:

\bcstart{D1} \textit{Support arranging, slicing and dicing~\cite{tory2021finding}, and semantically aligning sequences of charts depicting metrics}.

\bcstart{D2} \textit{Support the enrichment and annotation of metric chart sequences with situational context}.

\bcstart{D3} \textit{Provide context-aware recommendations of complementary metric charts that enrich a partially-constructed narrative sequence}.

\section{\remixtape}
\label{sec:remixtape}

\remixtape~is an application motivated in part by our formative study and design imperatives \textbf{D1---D3}, in part by prior research involving communicating around data in enterprise BI settings~\cite{brehmer2021jam,kandel2012enterprise,tory2021finding}, and in part by the BI industry's recent investments in applications and services focusing on metrics (see \cref{sec:rw:communication}).

Central to \remixtape's information architecture are collections of metrics as well as a hierarchical canvas of scenes and cards.

\bstart{Collections of metrics}
A BI metric is a univariate quantitative measure conventionally defined in reference to a temporal dimension and reported at a specified granularity (\eg day, week, month, quarter).
Metric values extend only as far back as to when the metric was defined by an analyst or data steward, and some metrics may no longer be accumulating values; consequently, different metrics will exhibit heterogeneous temporal domains (\ie the time span between a metric's starting value and its' most recent value), as well as heterogeneous granularities.
Anticipating this heterogeneity, \remixtape~is capable of ingesting multiple metric collections, with each metric containing a temporal attribute (\textcolor{rtblue}{\small{\faCalendar}}), at least one metric (\textcolor{rtgreen}{\small{\faHashtag}}), and (optionally) one or more categorical dimensions (\textcolor{rtblue}{\small{\faFont}}).

\bstart{Canvas, scenes, and cards}
\cref{fig:teaser} presents an overview of \remixtape's interface.
The input panel (\cref{fig:teaser}A) lists the available collections of metrics along with options to specify temporal granularities and filters.
Within the main canvas (\cref{fig:teaser}C), content is organized vertically in \textit{scenes}, which are in turn composed of one or more text or visualization \textit{cards} that are sequenced horizontally.

\remixtape~allows adding any number of scenes to the canvas, and these scenes can be reordered by dragging them within the canvas. 
It is similarly possible to append any number of visualization or text cards (\vizcards~or \textcards) to a scene, and cards can be reordered within a scene via dragging (\cref{sec:remixtape:semantic-align}).
\remixtape~provides a graphical summary of the canvas (\cref{fig:teaser}D), indicating each scene's metrics and its chronological coverage.

\textbf{\vizcard{s}} contain a line chart displaying one or more metrics. 
While initially empty, \vizcards~can be populated either by manually selecting metrics from the input panel or through system-generated recommendations (\cref{sec:remixtape:recommendations}).
With either approach, it is possible to specify a desired temporal granularity (day, month, year) and constrain the selected metrics' values to a specific time range (\cref{fig:teaser}A, bottom).
When combining metrics across datasets, \remixtape~implicitly finds common temporal granularities and collates time spans to render the appropriate line charts (\eg showing a gap in the line chart if there are non-overlapping time periods).

\textbf{\textcards} provide text input affordances for adding paragraph-length commentary within a scene (\textbf{D2}). 
Each \textcard~also provides affordances to add additional paragraphs as needed.

\subsection{Arranging and Semantically Aligning Views (D1)}
\label{sec:remixtape:semantic-align}

We offer several ways to arrange and coordinate \vizcards.

\begin{figure}[t!]
    \centering
    \includegraphics[width=\linewidth]{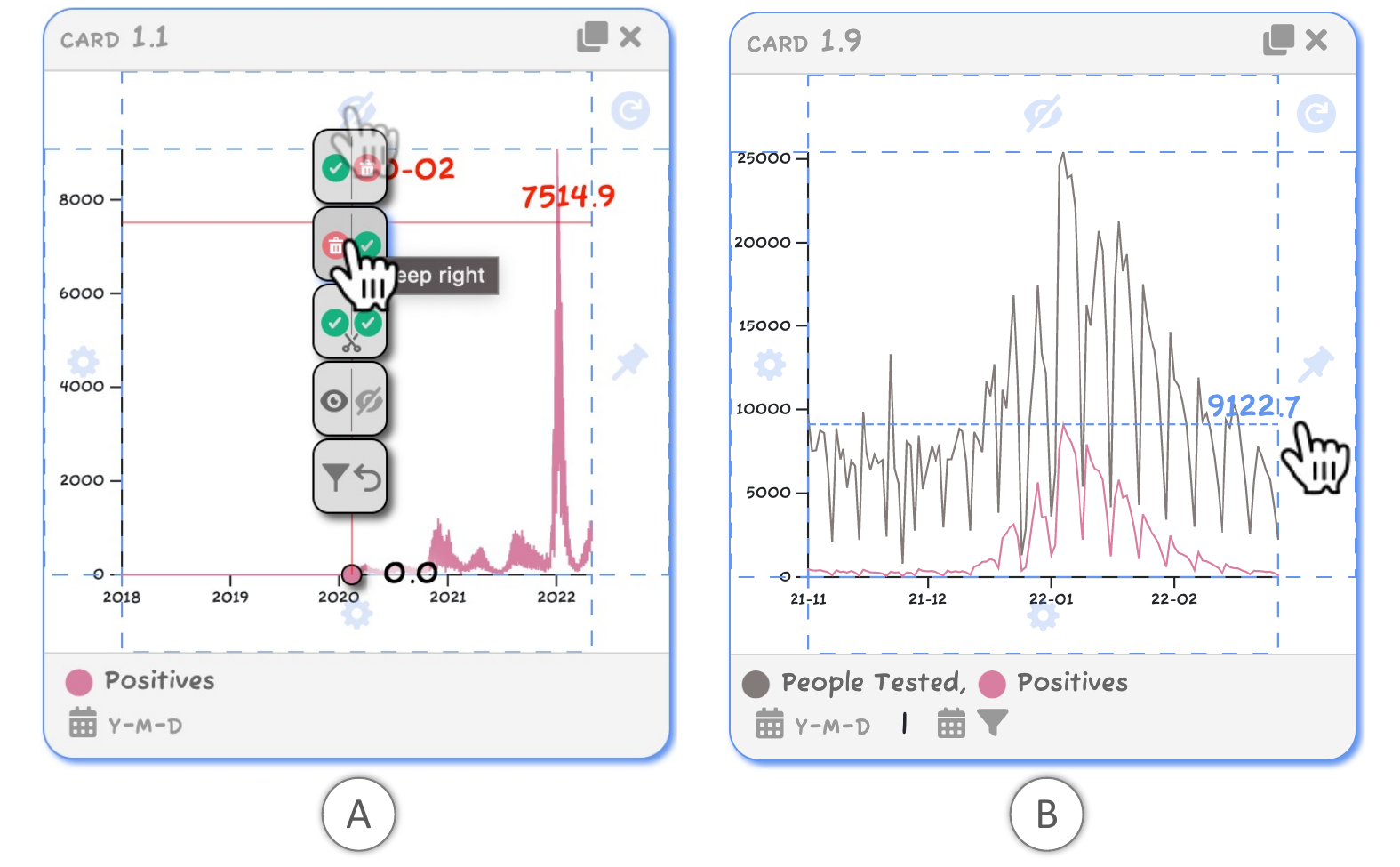}
    \caption{Examples of \vizcard~interaction affordances. (L) Clicking in the top margin provides options to filter, split, or obfuscate selected time spans. (R) Clicking in the right margin annotates the chart with a horizontal reference line to emphasize specific values.}
    \label{fig:scenario-2}
    \vspace{-1.5em}
\end{figure}

\lettrine[lines=4,findent=2mm,nindent=-0mm]{\includegraphics{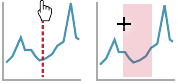}}{}\textbf{Slice and dice.}
Users may want to emphasize a specific time period by only showing the specific time span in a \vizcard. To this end, \remixtape~allows specifying temporal filters in the input panel before adding a \vizcard~to the canvas, as well as through interactions with a populated \vizcard. Specifically, clicking
in a \vizcard's top margin reveals controls for splitting the temporal domain of a chart (\cref{fig:scenario-2}L), including retaining data points either before or after the split point (\raisebox{-.5em}{\includegraphics[height=1.5em]{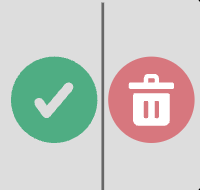}}, \raisebox{-.5em}{\includegraphics[height=1.5em]{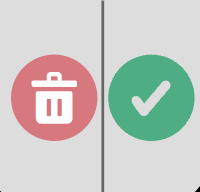}}), or moving data points after the split point to a new \vizcard~(\raisebox{-.5em}{\includegraphics[height=1.5em]{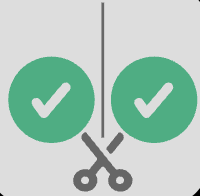}}).
Similarly, dragging parallel to the x-axis reveals controls to retain only the selected time span, or to exclude the selected span and retain the spans preceding and following it in two \vizcards.
Whenever a filter is applied to a \vizcard, we annotate it with icons ({\small{{\faCalendar*}~{\faFilter}}}) to indicate that the card only shows a subset of the data.

\begin{figure*}[t!]
    \centering
    \includegraphics[width=\textwidth]{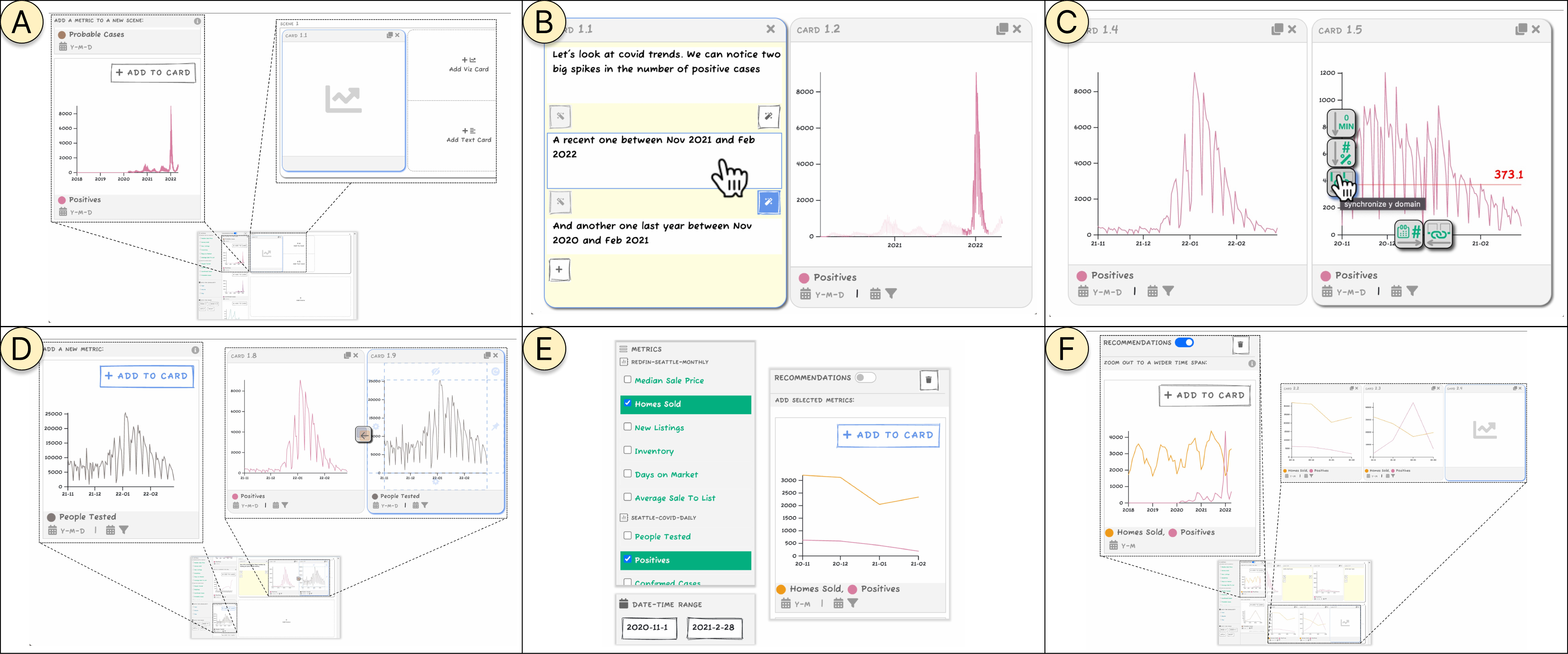}
    \caption{Using \remixtape~to construct a narrative about COVID-19 and housing sales. (A) Initial \vizcards~recommendations for a blank canvas. (B) Hovering over a \textcard~element referencing Nov 2021---Feb 2022 ephemerally highlights the corresponding time span in a juxtaposed \vizcard~(D2). (C) Options for Arranging and Semantically Aligning Views (D1). (D) A recommended \vizcard~of the \metric{People Tested} metric exhibits a trend similar to one seen in the preceding \vizcard~(\metric{Positives}). An option to merge (\raisebox{-.5em}{\includegraphics[height=1.5em]{figures/icons/merge.png}}) the two \vizcards~appears after adding the \metric{People Tested} \vizcard~to the scene. (E) A \vizcard~depicting  two metrics (\metric{Homes Sold} and \metric{Positives}) appears when manually selecting metrics from the metric list. (F) A recommendation of an overview \vizcard~appears when a scene contains \vizcards~with filtered time spans (D3).}
    \label{fig:scenario}
    \vspace{-1em}
\end{figure*}

\lettrine[lines=4,findent=2mm,nindent=-0mm]{\includegraphics{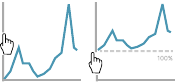}}{}\textbf{Truncate and relativize.}
Depending on the data context, one may want to show values along the Y-axis starting from the minimum value instead of zero (\eg in a line chart depicting temperature). 
Alternatively, in cases such as stock market metrics, it may be useful for a line chart to depict changes relative to a specific point in time. \remixtape~allows clicking on either axis within a \vizcard~to truncate and relativize the value domains of the chart (\cref{fig:scenario}C-right). Specifically,
the domain of the vertical axis can be toggled between the minimum and maximum values within the card or between zero and the maximum value (\raisebox{-.5em}{\includegraphics[height=1.5em]{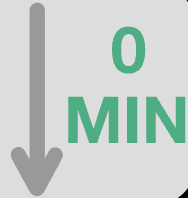}}, \raisebox{-.5em}{\includegraphics[height=1.5em]{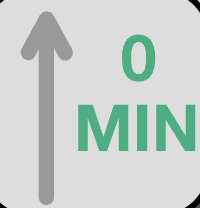}}).
The vertical axis can also be indexed to a percentage scale relative to the first value appearing in the card (\raisebox{-.5em}{\includegraphics[height=1.5em]{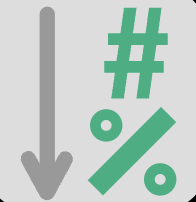}}).
A relative transformation can also be applied to the horizontal axis (\raisebox{-.5em}{\includegraphics[height=1.5em]{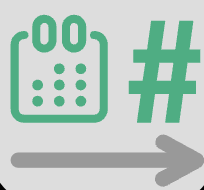}}), from absolute chronological dates to relative dates beginning from the first date appearing in the card.

\lettrine[lines=4,findent=2mm,nindent=-0.5mm]{\includegraphics{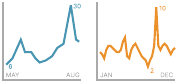}}{}
\textbf{Juxtapose and coordinate.}
Using consistent value domains across charts can enhance sequential narration and aid comprehension~\cite{qu2017keeping}.
When two \vizcards~are juxtaposed, \remixtape~can coordinate the axis domains of the right card with that of the preceding card by clicking on either axis in the right card and selecting the coordinate option  (\raisebox{-.5em}{\includegraphics[height=1.5em]{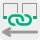}}, \raisebox{-.5em}{\includegraphics[height=1.5em]{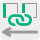}}, \cref{fig:scenario}C).

\lettrine[lines=4,findent=2mm,nindent=-0mm]{\includegraphics{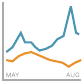}}{}\textbf{Merge and superimpose.}
While juxtaposing \vizcards~is one approach to comparing metrics, another approach is to combine them as a multi-series line chart in a single \vizcard.
Drawing upon the idea of semantic snapping~\cite{kristiansen2021semantic}, when two \vizcards~are juxtaposed, \remixtape~presents an option to merge them (\raisebox{-.5em}{\includegraphics[height=1.5em]{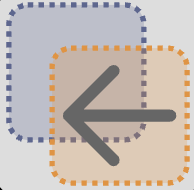}}) if they share a comparable value domain and time range (\cref{fig:scenario}D).
When the merge option is selected, \remixtape~replaces them with a new \vizcard~containing metrics from both cards, displaying the metrics' values for the overlapping time span between the two cards.

\subsection{Adding Situational Context (D2)}
\label{sec:remixtape:annotate}

\lettrine[lines=4,findent=2mm,nindent=-0mm]{\includegraphics{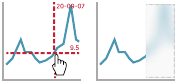}}{}\textbf{Annotate and obfuscate.}
Annotations can draw the audience's attention to specific time spans or values. \remixtape~supports adding both ephemeral or persistent annotations to \vizcards.
Ephemeral annotations are bound to the cursor position when mousing over a \vizcard, revealing horizontal and vertical reference lines and corresponding date and metric value text annotations.
A horizontal reference line and value annotation can also be persisted by clicking in a \vizcard's right margin (\cref{fig:scenario-2}R); clicking again clears persistent annotations.
Although additive annotations can reveal values, \remixtape~also provides support for dramatic subtractive reveals by adding a semi-opaque mask to obfuscate part of a chart, analogous to a fogged glass filter applied to obscure part of an image or a video (\raisebox{-.5em}{\includegraphics[height=1.5em]{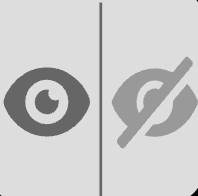}} in \cref{fig:scenario-2}L).
As with persistent annotations, clicking in the card's margin removes this obfuscatory mask, revealing the values beneath.

\lettrine[lines=4,findent=2mm,nindent=-0mm]{\includegraphics{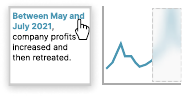}}{}\textbf{Coordinating \vizcards~\& \textcards.}
When arranged in a sequence, we act on the opportunity to coordinate a \textcard~with \vizcards~that precede or follow it. 
After commentary is added to a \textcard, each paragraph can be linked ({\small{\faMagic}}) to an adjacent \vizcard~(\cref{fig:scenario}B), invoking the SUTime library~\cite{chang2012sutime} for recognizing and normalizing time expressions in the paragraph.
Should a time expression intersect with the temporal domain of the linked \vizcard, subsequent mouse hover events on that paragraph will ephemerally emphasize the time span mentioned in the paragraph.
For instance, in \cref{fig:scenario}B, the \textcard~states 
\textit{``A recent one between Nov 2021 and Feb 2022.''}
Correspondingly, in the linked \vizcard~to the right, time periods before and after this span are de-emphasized until the mouse cursor departs from the paragraph.

\subsection{Recommending Charts from Context (D3)}
\label{sec:remixtape:recommendations}
Recommendations in \remixtape~are based on both the underlying data (\eg trends, correlations) and user actions (\eg metrics added to the canvas).
Specifically, \remixtape~presents two types of recommendations whenever a \vizcard~is added to the canvas: \textsc{Sequential} and \textsc{New Metric} recommendations.

For either category of recommendation, \remixtape~presents a maximum of five recommendations by default, giving users the option to explicitly request additional recommendations through a \textit{``See More\ldots''} button.
While recommendations are enabled by default, users can override them by manually specifying metrics, granularity, or temporal filters from the input panel (\cref{fig:teaser}A).
Users can also choose to disable recommendations altogether (\cref{fig:teaser}B).

\bstart{\textsc{Sequential} recommendations}
These recommendations retain the narrative's focus on a single metric while introducing new perspectives on that metric.
Specifically, when a \vizcard~is added to a partially-constructed scene, the system inspects the scene's content to generate two types of recommendations: \emph{drill-down} and \emph{overview}.

\istart{Drill-down recommendations}
\remixtape~recommends variations of charts already in a scene, adding temporal filters to drill down into prominent peaks and valleys.
The \textit{``Focus on a narrower time span''} recommendation in \cref{fig:teaser}B is an example of the former: upon parsing the trend in \metric{Homes Sold} values in the preceding card, \remixtape~recommends a card focusing on the drop in values in late 2021 and another focusing on the sudden peak in mid-2019, among others.
We employ a robust peak detection algorithm~\cite{brakel2014} that uses a moving average to detect peaks and valleys.
The algorithm uses three input parameters: 1) $lag$ controls the size of the moving window; 2) $threshold$ ($\lambda$) determines the number of standard deviations (\ie z--score); and 3) $influence$ controls the degree to which new data points will affect the moving average and standard deviation (a value in the range of $0-1$).
Based on empirical testing across datasets and temporal granularities, we set these parameters to 5 ($lag$), 3.5 ($threshold$), and 0.5 ($influence$), respectively.
For peak detection, at each time step $t$, a moving average $\mu_t$ and standard deviation $\sigma_t$ are calculated using data within the moving window.
A data point is then considered to be a peak if its value is greater than ${\mu_t} + {\lambda} * {\sigma_t}$ or a valley if its value is less than ${\mu_t} - {\lambda} * {\sigma_t}$.

\istart{Overview recommendations}
While \textit{drill-down} recommendations follow the overview~$\rightarrow$~detail narration pattern, it is possible to arrange a scene with \vizcards~that show only a small span of the time period for which metric data is available.
In these cases, there is an opportunity to provide missing context (\ie a detail~$\rightarrow$~overview transition).
Accordingly, \remixtape~recommends an overview chart spanning a wider time period (\eg{\cref{fig:scenario}F}).

\bstart{\textsc{New Metric} recommendations}
Metrics can contextualize one another, and while some of these relationships are foreseeable when constructing a narrative, we anticipate that there may be value in suggesting relationships serendipitously. 
To this end, \remixtape~recommends charts featuring metrics that are not yet on the canvas.

\istart{Recommendations for a partially constructed scene}
In cases where a new \vizcard~is added to a scene containing other populated \vizcards, \remixtape~recommends charts of other metrics that complement those already in the scene.
Specifically, \remixtape~iterates through available metrics and identifies those exhibiting trends that are similar to those in \vizcards~already in the scene.
For instance, in \cref{fig:teaser}, a chart of \metric{New Listings} is recommended because its trend is most similar to that of \metric{Homes Sold}, shown in the scene's penultimate card.
To detect similarity, the system computes the Pearson's correlation coefficient ($r$) between a metric's values and those displayed in the preceding \vizcard, recommending metrics with the highest $|r|$.

\istart{Recommendations for a new scene}
If other scenes containing populated \vizcards~are present in the canvas, \remixtape~suggests broadening the narrative by recommending charts of unused metrics that exhibit a high variance over time.

\istart{Recommendations for a blank canvas}
While we deem it unlikely that a user will not have a communication goal before using \remixtape, it nevertheless provides recommendations in a cold start scenario. 
In these instances, \remixtape~iterates through available metrics and recommends charts that exhibits a potentially interesting trend (\eg{\cref{fig:scenario}A}) by computing the coefficient of variance ($cv$) for metrics' values and recommending those with high variability.

\subsection{Usage Scenario}
\label{sec:remixing:scenario}

The following scenario reflects design imperatives \textbf{D1---D3} and is a composite of scenarios described by participants of the formative interview study. It centers on a fictional real estate professional named Sarah who will use \remixtape~to prepare a presentation \revision[]{for her colleagues} about the dynamics of COVID-19 and the housing market in Seattle, Washington.
Her organization's data stewards have defined a collection of metrics tracking Seattle's daily COVID-19 statistics since 2020, as well as another collection tracking Seattle's monthly real estate listings since 2018.
Her narrative will reference a few metrics spanning both collections.

Sarah begins by adding a scene and a \vizcard~to the canvas.
She decides that spikes in the number of COVID-19 \metric{Positives} (\cref{fig:scenario}A) will establish a familiar context with her audience, so she populates her first \vizcard~with this metric.

Since \metric{Positives} does not have values prior to February 2020, Sarah filters the \vizcard~to focus on what has transpired since then (\cref{fig:scenario-2}L).
The cards show two prominent spikes in the data: a massive spike at the end of 2021 and a smaller one at the end of 2020.
She fills a \textcard~to draw attention to these spikes in and links these comments to the \vizcard~(\cref{fig:scenario}B); this will highlight these spans when interacting with the \textcard~(\textbf{D2}).

To draw attention to the two spikes, Sarah creates two duplicates of the \vizcard~showing \metric{Positives} and filters them to time periods of the two spikes.
Noticing that their vertical axes are misaligned, she coordinates them (\cref{fig:scenario}C, \textbf{D1}) to draw attention to the fact that the recent spike coinciding with New Year's 2022 is almost eight times greater than that of the earlier spike, which coincided with the 2020 Thanksgiving holiday.

To add context to the two spikes, Sarah duplicates the \vizcard~showing the New Year's \metric{Positives} spike and considers \remixtape's \textit{new metric recommendations} (\textbf{D3}).
She adds the \metric{People Tested} recommendation to her scene (\cref{fig:scenario}D).
She then merges the two cards and similarly creates another card depicting the dynamics of \metric{Positives} and \metric{People Tested} during the Thanksgiving 2020 spike.
Next, she adds a \textcard~to indicate that the rate of positives was much higher during the recent spike relative to during the previous spike.
To emphasize this drastic difference in the number of \metric{Positives}, she adds reference line annotations orthogonal to the vertical axes (\cref{fig:scenario-2}\revision[B]{-right}, \textbf{D2}).

Next, Sarah expands her narrative to include housing metrics.
She adds a new scene and selects the \metric{Homes Sold} and \metric{Positives} metrics from the metric list in the input panel.
To mirror the recent COVID spike featured in the first scene, she explicitly specifies November 2021 to February 2022 as her time period of interest.
\remixtape~then presents a multi-series line chart that includes the two metrics, automatically aggregating the \metric{Positives} to match the monthly granularity of the \metric{Homes Sold} metric (\cref{fig:scenario}E, \textbf{D1}).
She adds this chart to the scene and similarly creates a second \vizcard~spanning the period corresponding to the earlier spike.

Sarah's narrative sequence now includes two \vizcard{s} in the second scene that depict a dip in \metric{Homes Sold} during the last two months of both 2020 and 2021, periods coinciding with rises in \metric{Positives}.
As she adds a new card to contextualize these periods, \remixtape~observes that the cards in this scene show a narrow time range; accordingly, it provides an \textit{overview recommendation} spanning the entire available time range for the housing metrics, from January 2018 until April 2022 (\cref{fig:scenario}F, \textbf{D3}).
Now her narrative communicates that seasonal drops in \metric{Homes Sold} in the fall were common even before the pandemic.
She then appends a \vizcard~to the scene showing the values for the months before February 2020 and adds an explanation in a new \textcard.

To expand the narrative to other real estate metrics beyond sales, Sarah adds a new \vizcard~showing only the \metric{Home Sales} metric and peruses the \textit{new metric recommendations} that she could add.
The \metric{New Listings} exhibits a similar pattern to sales (\cref{fig:teaser}, \textbf{D3}), so she creates a merged \vizcard~composed of \metric{Home Sales} and \metric{New Listings} (similar to \cref{fig:scenario}D).

Satisfied with her two scenes that combine metrics from different sources that span partially overlapping time periods, Sarah feels confident in communicating her narrative.
Later, in a meeting with her colleagues, she \revision[screen-shares]{toggles} \remixtape's presentation mode as she presents her interpretation of the dynamics of Seattle's housing market before and during the COVID-19 pandemic \revision[]{(the toggle is located in the header at the top right of \cref{fig:teaser}, and the presentation mode is} \revision[see]{demonstrated in the supplemental} video).

\section{User Study}
\label{sec:evaluation}

We observed enterprise professional participants perform two activities with \remixtape~reflecting \textbf{D1}---\textbf{D3}. 
We adopted a methodology of reproduction and open-ended construction activities established in prior evaluations of interactive construction applications in the visualization literature~\cite{Amini2018DDS,ren2018reflecting}.
Respecting our participants' valuable time, this methodology also allowed to elicit candid feedback pertaining to \remixtape's potential utility, as well as ideas for extending and integrating it into existing workflows.
Note that the goal of this study was not to generate objective evidence of utility or usability, but rather to present a qualitative reflection on \remixtape's plausibility~\cite{meyer2019criteria} informed by multiple subjective expert perspectives.

\subsection{Participants}
\label{sec:evaluation:participants}

We recruited six professionals (P1---P6) distributed across North America and Europe who routinely communicate narratives concerning data with their colleagues, stakeholders, and customers.
They responded to our call for participation that we disseminated through mailing lists and Slack channels at a multinational software corporation and its subsidiary companies.
None of these professionals had participated in our initial study described in \cref{sec:remixing}.
We screened respondents based on their form responses to questions regarding their prior experience with respect to communicating narratives grounded in data, with a focus on time-series metrics.
All participants reported using business intelligence (BI) assets created by others, such as dashboards, metrics, or charts. 
Half of the participants regularly use or refer to content from five or fewer dashboards in the course of their work, while the other half regularly use content from more than ten.
While they shared interests and experiences with respect to communicating narratives about metrics, they varied in terms of gender, years of professional experience, and job role (three held management roles, three held customer-facing roles).

\subsection{Procedure}
\label{sec:evaluation:procedure}

We conducted study sessions remotely via video conferencing.
One author moderated the sessions while the other authors took notes.
We recorded and transcribed each session, which included screen capture recordings of  \remixtape~usage.
Each session lasted between 75 and 90 minutes and included four phases, with the majority of this time split between the two hands-on activities.

\bfstart{Introduction}.
We began each session with a brief interview to understand current workflows with respect to  metrics in communication contexts and any associated challenges.
Next, we played a three minute-long video tutorial of \remixtape~to familiarize participants with its major features.
We then gave participants an opportunity to ask clarifying questions and provided them with a user interface reference document that they could refer to during the activities.

\bfstart{Activity 1: Canvas reproduction}.
The first activity served as hands-on training as well as a preliminary assessment of \remixtape's usability and learnability.
We asked participants to screen-share their web browser and proceed to a URL where we had deployed an instance of \remixtape~populated with two collections of metrics: the monthly Seattle housing metrics introduced in \cref{sec:remixing:scenario} and a set of daily Seattle weather metrics, with both collections of metrics spanning January 2018 to June 2022.
Next, we provided participants with a screenshot of a \remixtape~canvas populated with two scenes, each containing two \vizcards~and two \textcards: a sequence representing a partial narrative about the dynamics of Seattle's housing and weather.
The activity required participants to reproduce the canvas as illustrated in the screenshot beginning from an empty canvas, which involved selecting metrics and specifying their temporal granularities, arranging and coordinating \vizcards~via filtering, relativizing, and merging them (\textbf{D1}), as well as annotating \vizcards~and adding \textcards~(\textbf{D2}).
This activity allowed flexibility in terms of the sequences of interactions leading to a correctly reproduced canvas. 
The only restriction that we imposed was the avoidance of the recommendation feature, which we set aside for the second activity.
Finally, we encouraged participants to think aloud and indicate when they encountered difficulties with the interface so that we could resolve any confusion.

\bfstart{Activity 2: Open-ended construction}.
The second activity was a more free-form construction scenario~\cite{ren2018reflecting}.
While continuing to screen-share their web browser, we directed participants to another URL where we had deployed another instance of \remixtape~featuring the COVID-19 and housing metrics introduced in \cref{sec:remixing:scenario}.
Unlike in the first activity, we pre-populated the canvas with a scene containing two \textcards~and three \vizcards~that served to introduce a narrative about two major spikes in COVID-19 cases.
This activity required that participants either extend this initial narrative or establish their own narrative about the potential dynamics between these two sets of metrics.
Also unlike the first activity, we encouraged participants to make use of the recommendations feature as they completed the activity (\textbf{D3}).
Otherwise, we similarly encouraged participants to think aloud as they populated the canvas with \vizcards~and \textcards.

\bfstart{Debrief}.
Finally, we asked participants to comment on how \remixtape~could be extended or integrated into their workflows.  

\subsection{Findings}
\label{sec:evaluation:findings}

The lead author performed an initial thematic analysis of our observations in reference to design imperatives \textbf{D1}---\textbf{D3}, later engaging the other authors in discussion and focused analysis.

\bstart{Arranging sequences of cards~(D1)} 
Every participant remarked upon the \revision[]{unfamiliar} scene- and card-based organization of \remixtape's canvas, \revision[reflecting a common realization]{acknowledging} that \remixtape~is neither a conventional BI dashboard interface nor a slide presentation tool. 
Anticipating the former, P2 described how the introduction of scenes and cards had them \textit{``feeling a little lost''}, while by anticipating the latter, P5 felt similarly lost without the interface conventions of slide presentation tools.
Looking past these differences, P6 appreciated our decision to place equal weight on \vizcards~and \textcards, allowing people to place them in any order in the canvas:     
\textit{``The text just makes it more approachable and gives people the confidence to be really sure what they're supposed to be taking away.''} 

While the juxtaposition of \textcards~and \vizcards~could help to keep teams \textit{`on the same page'} [P6], P1 suggested adding a way to toggle between the display of a single \vizcard~and a juxtaposed display of multiple cards when using \remixtape~as a synchronous presentation tool.
Similarly, P3 suggested collapsing scenes so that they could be revealed on demand, such as scenes prepared in anticipation of audience questions during a live presentation.
Going a step further, P4 saw the selective reveal of charts as being critical for persuading the audience and guiding attention:
\textit{``if the idea is to persuade, then that would be the best way to do it because I can't control what} [the audience is] \textit{viewing before I get there.''}
P4 also appreciated the affordances for arranging and coordinating \vizcards~in \remixtape~and envisioned selectively revealing content at finer levels of specificity, such as revealing individual metrics appearing in a multi-series \vizcard.

P4 also anticipated situations where a presenter might limit the scope of where and how their audience could interact with charts appearing on a shared canvas, describing how an audience could \textit{``click on a particular \textcard~that exposes a \vizcard~once I've unlocked access to that, because I'm at that part of the scene,''} likening the process to the selective muting and unmuting of instrument channels in live musical performance software.

Scene and card hierarchies elicited several alternative suggestions with respect to arranging cards within a \remixtape~canvas. 
First, P2 questioned why we limited the interface to a two-level hierarchy and suggested the possibility of nesting cards or creating those with a focal chart with multiple peripheral charts, which could be useful for displaying a single aggregate metric alongside a categorical breakdown of values contributing to that aggregate.
Second, P3 appreciated the logical separation of cards into scenes but found the vertical ordering of scenes to be limiting, describing use cases where two scenes each containing two cards are placed to be horizontally adjacent.
P5 provided a similar suggestion of \textit{``a vector canvas where you can just layer the cards} [and] \textit{freely place them} [\ldots] \textit{or toggle between floating and tiled, hierarchical and linear.''} 
A free-form canvas environment could reinforce the informality of some conversations around data, with P5 predicting that it would \textit{``be really good for a rehearsal or dry run}\textit{''} of a presentation.

To P2, the seemingly informal process of generating \vizcards~and arranging them on the canvas is one that could accelerate individual storyboarding workflows:
\textit{``}{\vizcards} \textit{are very cheap and I can spend the whole day creating} \textit{and leaving them until I find something interesting, but I cannot do that in} [existing tools]\textit{.''}
However, when considering different audiences, contexts of use, and the lifespan of communicative artifacts, expectations regarding information architecture and required functionality may shift. 
In particular, to support ephemeral and synchronous conversations around metrics with colleagues, P2 remarked that the canvas / scene structure along with functionality for annotating and semantically aligning cards may not be warranted. 
In contrast, such functionality would be important to retain for longer-living communication artifacts consumed by audiences asynchronously.

\bstart{Slicing and dicing cards (\textbf{D1}) and annotating them (\textbf{D2})}
Participants' reactions to the affordances for coordinating and arranging cards were mixed.
While P5 felt that the \textit{``The design mode works very fluidly''}, P2, P3, and P6 needed reminders of interactions invoked from a \vizcard's margins (\cref{fig:scenario}C, \cref{fig:scenario-2}), often right-clicking on the chart area to invoke a context menu, seeking options for filtering and adding annotations.
Furthermore, when performing filter operations like card splitting, P6 failed to realize that a filter was applied to the \vizcard~(\ie they did not notice the {\small{{\faCalendar*}~{\faFilter}}} status icons on the card).
However, these concerns subsided as participants became more accustomed to the interactions during the activities.
Citing tutorial design in video games, P5 suggested including in-situ onboarding features to aid discoverability, as these could reduce the initial learning curve; these features could manifest either as a stepped walkthrough of portions of the interface or via animated tooltips revealed when placing the cursor over specific elements.

All of the participants appreciated the temporal comparison \remixtape~supports by allowing them to spatially juxtapose, semantically align, and especially merge metrics into a single \vizcard.
However, it was not always clear to participants whether or how they could semantically align or merge two \vizcards.
For instance, four participants tried to merge \vizcards~that exhibited different time granularities or time spans; they wanted to align temporal patterns at different scales or across non-overlapping time periods, such as to highlight seasonal patterns.
Upon realizing that merging \vizcards~was not possible in these cases, participants asked \textit{``How would I allocate for this lag [between two metrics]? Can I offset [one metric]?''} [P4] or if the system could \textit{``union them in a way''} [P5].
Such points of confusion with respect to semantic alignment interactions suggest that the system needs to provide more clarity, not only with respect to when \vizcards~can be merged, but also to explain cases where a merge is not feasible.

Irrespective of their experience with the card coordination and annotation interactions in \remixtape's design mode, all participants remarked positively on the ability to directly interact with content in the presentation mode.
For example, P6 appreciated the \textit{`dynamic nature'} of the presentation mode, providing affordances to tell \textit{``more of a story than a static screenshot.''}    
P6 also commented that linking text with charts dynamically would support synchronous communication with audiences. However, P4 expressed a desire for additional control over text positioning and appearance (\eg through tooltips or the progressive reveal of \textcards~in presentation mode).
Finally, P4 and P6 alluded to both presenters and audiences having the ability to interact with cards in \remixtape's present\revision[]{ation} mode, which suggests \revision[additional investment into]{that \remixtape~could be extended with} real-time multi-user collaboration \revision[support]{functionality}.

\bstart{Recommending contextually relevant \vizcards~(\textbf{D3})}
The value proposition of recommendations varied based on prior knowledge of a dataset and whether one already has a comprehensive narrative in mind prior to using the tool, with P6 stating:
\textit{``I like the recommendation engine for people like me} [who] \textit{are more business users. I don't really know the story I'm trying to tell, and I don't know tools as intimately as analysts do.''} 
P2 referred to recommendations as their \textit{`favorite feature'} after seeing \remixtape~suggest charts across a collection of metrics spanning various temporal granularities.
Participants also saw the potential value in recommendations as a form of guidance when initially forming a narrative, or as a source of contextually-relevant ideas during narrative construction.
For instance, P5 stated that \textit{``the recommendations feature is interesting in that it's really looking at what else is currently in the story.''}

In contrast, P3 questioned the role of recommendations when working with data that they were \revision[more]{already} familiar with.
\noindent{Given that \remixtape~\textit{``feels more like a presentation tool''} and does not offer rich data exploration affordances (noting that \textit{``this experience doesn't scream data exploration to me''}), its recommendations may have limited value for those who already have a narrative in mind after conducting analysis with other tools.}

Participants also commented on the types of recommendations (\cref{sec:remixtape:recommendations}).
P2 explained that in some instances, they care more about a particular metric and they will \textit{``dig deep into it and explore it''}, suggesting the value of \textsc{Sequential} recommendations and particularly those of the \textit{drill-down} variety. 
In other instances, they might care more about the relationship between metrics, suggesting the value of \textsc{New Metric} recommendations.
Currently, the organization of scenes on the canvas affect the type of recommendations shown, but P2's comments suggest an ability to take more explicit control over the type of recommendations generated by \remixtape.
This sentiment was echoed in another session, in which P3 sought a way to \textit{`pre-filter'} the type of recommendations that \remixtape~would generate. 

One potential concern pertains to the specificity of recommendations, as voiced by P4:    \textit{``Is it almost unethical to say `Hey, recommendation model, I'm trying to prove a correlation between `this' and `this'?} [\ldots] \textit{Will} [\remixtape] \textit{populate recommendations that support my hypothesis?'} 
Setting aside the issue of whether such a degree of specification is possible or ethical, P4 also suggested a potential mitigating strategy that could be applied to \vizcards: the annotation of individual \vizcards~to indicate that they were either manually specified by the user or generated via \remixtape's recommendation engine.
These annotations could serve to communicate the provenance of a narrative and establish a degree of trust in both the user who constructed the canvas as well as in the system.

\section{Discussion}
\label{sec:discussion}
We critically reflect on our design imperatives and consider alternatives suggested by our study findings, as well as on aspects of \remixtape~that could apply beyond metrics and BI.


\subsection{Reflecting on Study Findings \& Key Takeaways}
\label{sec:discussion:interpretation}

\bstart{Deviating from dashboards, slides, and galleries}
In our formative interview study (\cref{sec:remixing}), interviewees described a need to present and contrast metrics, echoing a similar need articulated in an earlier interview study by Brehmer and Kosara~\cite{brehmer2021jam}. 
In their study, they documented the use case of revealing metric values in such a way that allows a presenter to make a time-over-time contrast, such as juxtaposing or superimposing an organization's most recent quarterly sales performance against the same quarter of the previous year.
While such a contrast could, in some instances, be shown within a single BI dashboard, Brehmer and Kosara's interviewees remarked that dashboard applications were not well-suited for narrative presentations, and that attempting to use them in such a way gave the impression of a tool demonstration rather than as a medium for storytelling. 
With the advent of unified metrics layers (\eg~\cite{powerbimetrics,tableaupulse}), metrics no longer need to be housed in individual dashboards, but can rather be aggregated into collections displayed as galleries~\cite{tableaucollections}; however, such an interface is still better suited for browsing and monitoring.  
If dashboards and galleries are not effective presentation tools, people turn to slide presentations. 
We know from Tory~\etal~\cite{tory2021finding} that BI presentations often involve \textit{`slicing and dicing'} charts and other BI assets and combining them into linear slide presentations, often as static screenshots.
\remixtape's scrollable storyboard canvas therefore represents an alternative to dashboards, galleries, and slide presentations for preparing and delivering narratives about metrics.
However, as reported in \cref{sec:evaluation:findings}, our alternative organizing structure was disorienting for several participants, suggesting either a new onboarding experience or an ability to toggle to \revision[and from] more familiar \revision[]{or simplified} modes of content organization. 
\textbf{Takeaway}: \textit{Exercise caution when proposing alternative narrative BI experiences that deviate from the expected conventions of dashboard or slide presentation interfaces.}

\bstart{Extending the scene + card canvas for narratives about metrics}
We did not reject the conventions of slide presentation software altogether: cards in \remixtape~are analogous to individual slides, with each slide containing either an interactive line chart or an interactive block of text.
While our study participants generally appreciated the simplicity of cards, the equal treatment of text and visualization, and the chunking of cards into scenes, their comments also suggest that we could relax our imposition of a linear order of cards.
To this end, we could assign a level of importance to individual metrics.
For example, in Elshehaly~\etal's QualDash application~\cite{elshehaly2020qualdash}, a card-based dashboard specification involved assigning healthcare quality metrics to either a primary or secondary importance; while hidden by default, secondary supporting metrics in QualDash were revealed in an expanded card of a primary metric, an approach that could satisfy
P3 and P4's related suggestions (\cref{sec:evaluation:findings}).
The flexible \revision[organizing]{} structure of comic strips~\cite{bach2018design,zhao2019understanding} is another \revision[alternative]{(and possibly simpler)} organizing scheme for metrics on a canvas. 
This approach is not incompatible with assigning importance levels to metrics, in that panels associated with secondary-class metrics could be smaller or less visually prominent, using other conventions for showing metrics beyond line charts, such as sparklines~\cite{tufte2006} or BANs~\cite{wexler2017}. 
\textbf{Takeaway}: \textit{Consider scene + card-based presentations about metrics with cards of varying visibility and prominence.}

\bstart{Rethinking content recommendations for BI presentations}
Given our study findings, we acknowledge a potential mismatch between a desire for serendipitous content recommendations while formulating a narrative and \remixtape's hierarchical scene + card-based canvas. 
These recommendations might be better suited when initially placing cards on an unbounded whiteboard canvas, as we suggested to participants of our formative interview study. 
Moreover, content recommendations might be altogether unnecessary if one already has a clearly-formulated narrative.
Alternatively, recommendations could be offered \textit{after} a narrative sequence is drafted:
\remixtape~could recommend alternative card sequences exhibiting different rhetoric or argument structures~\cite{hullman2013deeper,lan2021understanding}, or suggest perception-based intermediary \vizcards~that make for smoother transitions between points in a narrative~\cite{kim2017graphscape,lin2020dziban}.

Another opportunity is recommending further synchronization between visualization and text.
Currently, \remixtape~provides the ability for users to compose text and link this text to spatially adjacent \vizcards, interactively highlighting portions of the latter when the text is brought into focus.
Beyond this synchronization, there is an opportunity to leverage so-called \textit{`auto-insight'} features from recent work on natural language generation for describing charts~\cite{wang2019datashot,srinivasan2018augmenting,cui2019datasite}; for example, when adding a \textcard~to the canvas, \remixtape~could generate an initial description of its adjacent \vizcards, which the user could opt either to retain, edit, or extend.
The converse is also possible~\cite{bendeck2024slopeseeker}: for instance, when composing a \textcard~with nouns (\eg~\textit{`spike'}), adjectives (\eg~\textit{`dramatic'}), verbs (\eg~\textit{`falling'}), or adverbs (\eg~\textit{`gradually'}) that are typically associated with line charts, the system could recommend corresponding highlights and annotations on adjacent \vizcards, or it could recommend the addition of other \vizcards~that depict similar patterns to those described in natural language.
\textbf{Takeaway}: \textit{Narrow the scope of content recommendations (e.g., alternative sequential structures, tighter text-chart alignment); otherwise, recommendation features give an impression that the application is intended for exploratory data analysis rather than for communicating narratives}.

\subsection{Limitations \& Future Work}

\bstart{Beyond BI and metrics}
\revision[]{Our focus on metrics reflects the broader BI software community's recent investments in cloud infrastructure for shared metric repositories~\cite{powerbimetrics,tableaumetrics,databrainmetrics,dbtmetrics,transformmetrics,tableaupulse}.}
Although we situate \remixtape~within an enterprise BI context~\cite{brehmer2021jam,kandel2012enterprise,tory2021finding}, aspects of this project may be applicable in other domains where there is a need to construct and present narratives around data. 
STEM education or policy communication are two that come to mind, particularly as public and civic data portals provide opportunities to collect metrics originating from different institutions and agencies.
\revision[Our focus on metrics also reflects the broader BI software community's recent investments in cloud infrastructure for shared metric repositories. Enterprise conversations around data continue to make use of other BI assets, from dashboards and tables to spatial data displayed on maps and non-spatial data represented by other chart design conventions.]{}

\bstart{A design space for visualization remixing}
The name \remixtape~is a portmanteau of \textit{remix} and \textit{mixtape}, with both terms being allusions to music production and dissemination. Additionally, the latter imparts a nuance of continuity: a sequential presentation of re-contextualized content. 
While the immediate purpose of \remixtape~is to support presentations about sequences of metrics, it nevertheless reflects a broader desire to \textit{slice and dice} BI assets and enrich them with situational context, one emanating both from our own formative study (\cref{sec:remixing}) and from interviews with enterprise professionals in earlier work~\cite{brehmer2021jam,tory2021finding}.
We also acknowledge that a need to remix the artifacts of knowledge work transcends the BI domain~\cite{feuston2018social}.
\remixtape~therefore represents a point in a design space for \textit{visualization remixing}, and we call upon the research community to define a conceptual foundation for this design space and to differentiate it from remixing in other contexts.

\subsection{Additional Evaluation Opportunities}

\bstart{Additional evaluation opportunities}
Selecting an appropriate evaluation methodology for open-ended narrative visualization construction applications like \remixtape~can be challenging~\cite{ren2018reflecting}, particularly with many constructs to consider (\eg~learnability, usability, workflow integration, expressiveness~\cite{Amini2018DDS}).
Additionally, comparative studies are often logistically infeasible for these applications due to the confounding differences in feature sets and interaction paradigms~\cite{satyanarayan2019critical}. 
Redesigning aspects of \remixtape~following the takeaway lessons of \cref{sec:discussion:interpretation} and conducting a longitudinal post-deployment study centered around collections of metrics already familiar to participants will allow us to better understand \remixtape's potential for communicating narratives about metrics and the role of these narratives in decision making~\cite{dimara2021unmet}.
\section{Conclusion}
\label{sec:conclusion}

The relationship between visualization research and business intelligence (BI) software has evolved over time, from the former's contributions to the conventions of interactive self-service visual analysis two decades ago~\cite{stolte2002polaris} to the research community's renewed interest in BI dashboards since the end of the 2010s~\cite{bach:2022dashboardDesignPattern,sarikaya2018we,tory2021finding,walchshofer2023transitioning}, which in turn yielded new application-based research on approaches to dashboard construction (\eg~\cite{lin2022dminer,pandey2022medley}). 
As many dashboards could be simply described as containers for time-oriented metrics, our formative interview study highlights the frustration that enterprise professionals feel when attempting to collect, recontextualize, and present metrics from disparate dashboards. 
\remixtape~is a response both to these frustrations and to the emergence of shared metrics platforms within BI ecosystems~\cite{epperson2023declarative,tableaumetrics,powerbimetrics,databrainmetrics,dbtmetrics,transformmetrics,tableaupulse}, wherein sets of metrics may be associated with heterogeneous data sources and time frames.
Lastly, \remixtape~represents a point in the nascent design space of visualization remixing, one that transcends time-oriented metrics and BI contexts, a design space that should be systematically explored and defined in future research.


\acknowledgments{%
The authors conducted this research while at Tableau; we thank Zach Morrissey, Marlen Promann, Carolyn Tweedy.
}

\bibliographystyle{abbrv-doi-hyperref}

\bibliography{main}

\end{document}